\documentclass[RNAAS]{aastex62}

\begin{document}

\title{
Distance and Tangential Velocity of the Main Ionizing Star in the North America/Pelican Nebulae with Gaia EDR3
}

\correspondingauthor{Michael A. Kuhn}
\email{mkuhn@astro.caltech.edu}

\author{Michael A. Kuhn}
\affiliation{California Institute of Technology, Pasadena, CA 91125, USA}

\author{Lynne A. Hillenbrand}
\affiliation{California Institute of Technology, Pasadena, CA 91125, USA}

\begin{abstract}
The Bajamar Star is an early O star that ionizes the North America/Pelican Nebulae. In projection, it is near the geometric center of the H\,{\sc ii} region, but appears to lie outside any of the main stellar subgroups. Furthermore, in Gaia DR2, there were slight discrepancies between this star and the rest of the system in parallax (2$\sigma$ larger) and relative tangential velocity ($\sim$6~km~s$^{-1}$). Using Gaia EDR3, we find that the parallax discrepancy has disappeared, but the velocity difference remains. These results are consistent with the star having escaped from a subgroup.  
\end{abstract}

\section{} 

Early O-type stars are expected to have significant influences on star-forming complexes due to the disruptive effect of their strong radiation and winds on the molecular clouds and their dominant effects on stellar $n$-body dynamics \citep[e.g.,][]{Dale2015,Oh2016}.

The North America/Pelican Nebula Complex (hereafter NAP) is a star-forming region at a distance of $\sim$800~pc \citep{Kuhn2020,Zucker2020} containing one of the nearest O3.5--O5 type stars \citep[the Bajamar Star = 2MASS J20555125+4352246;][]{Comeron2005,MA2016}. This source is projected near the geometric center of the H\,{\sc ii} region \citep{Bally1980,Comeron2005}. However, there are two challenges in explaining its relation to the star-forming complex. 

First, the Bajamar Star is outside the known subgroups of YSOs \citep{Kuhn2020,Fang2020}. Using Gaia DR2, \citet{Kuhn2020} provide a possible explanation. The tangential $\sim$6~km~s$^{-1}$ velocity of the O-star in the reference frame of the association, which is faster than most other members, implies that it could have escaped from one of the groups $\sim$1.5~Myr ago. 

Second, the Gaia DR2 parallax of the Bajamar Star ($\varpi_\mathrm{DR2}=1.47\pm0.08$~mas), taken at face value, would place it in front of most other stars in the complex ($\varpi_\mathrm{DR2}\approx1.2$~mas). This includes a group of stars embedded in the ``Gulf of Mexico'' (L935) cloud; however, this cloud also appears to obscure the Bajamar Star ($A_V\sim9.6$~mag). Because the geometric requirement that the obscuring cloud be between us and the Bajamar Star outweighs the parallax discrepancy, \citet{Kuhn2020} attribute the difference to a $\sim$2$\sigma$ statistical error in the star's DR2 parallax measurement. 

Gaia's early data release 3 \citep[EDR3;][]{Gaia2020} provides higher precision astrometry relative to DR2 and improved treatment of systematic effects \citep{Lindegren2020_as}. Furthermore, \citet{Lindegren2020_zp} provide parallax zero-point estimates as a function of ecliptic latitude, magnitude, and spectral shape. These corrections should assist in comparing the O star to other stars in the complex, given that the Bajamar Star ($G=10.5$~mag and $BP-RP=3.9$) is much brighter and redder than most other members. In EDR3 the Bajamar Star has a 5-parameter solution with a small, but statistically significant astrometric excess noises of $\sim$0.3~mas, and a reduced unit weight error of $\sim$1, implying no known astrometric problems.  

For comparison, we use the same sample of 395 NAP members as \citet{Kuhn2020}. We computed their corrected EDR3 parallaxes, using zero points ranging from $-0.067$ to $-0.012$~mas  (median of $-0.043$~mas). The zero point for the Bajamar Star is $-0.060$~mas, yielding a corrected parallax of $\varpi = 1.327\pm0.030$~mas. 

\begin{figure}[t]
\begin{center}
\includegraphics[width=0.48\textwidth]{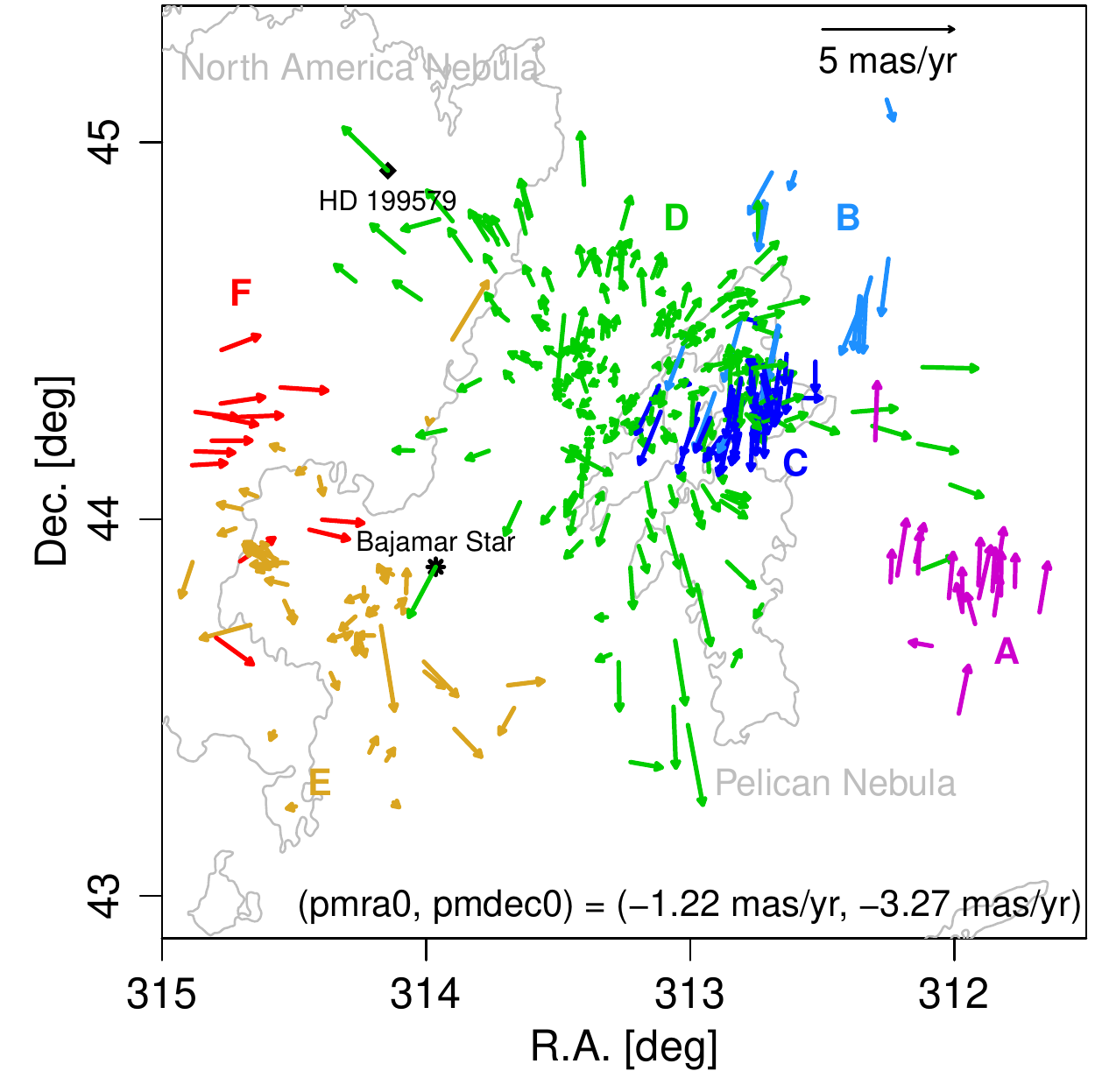}
\includegraphics[width=0.48\textwidth]{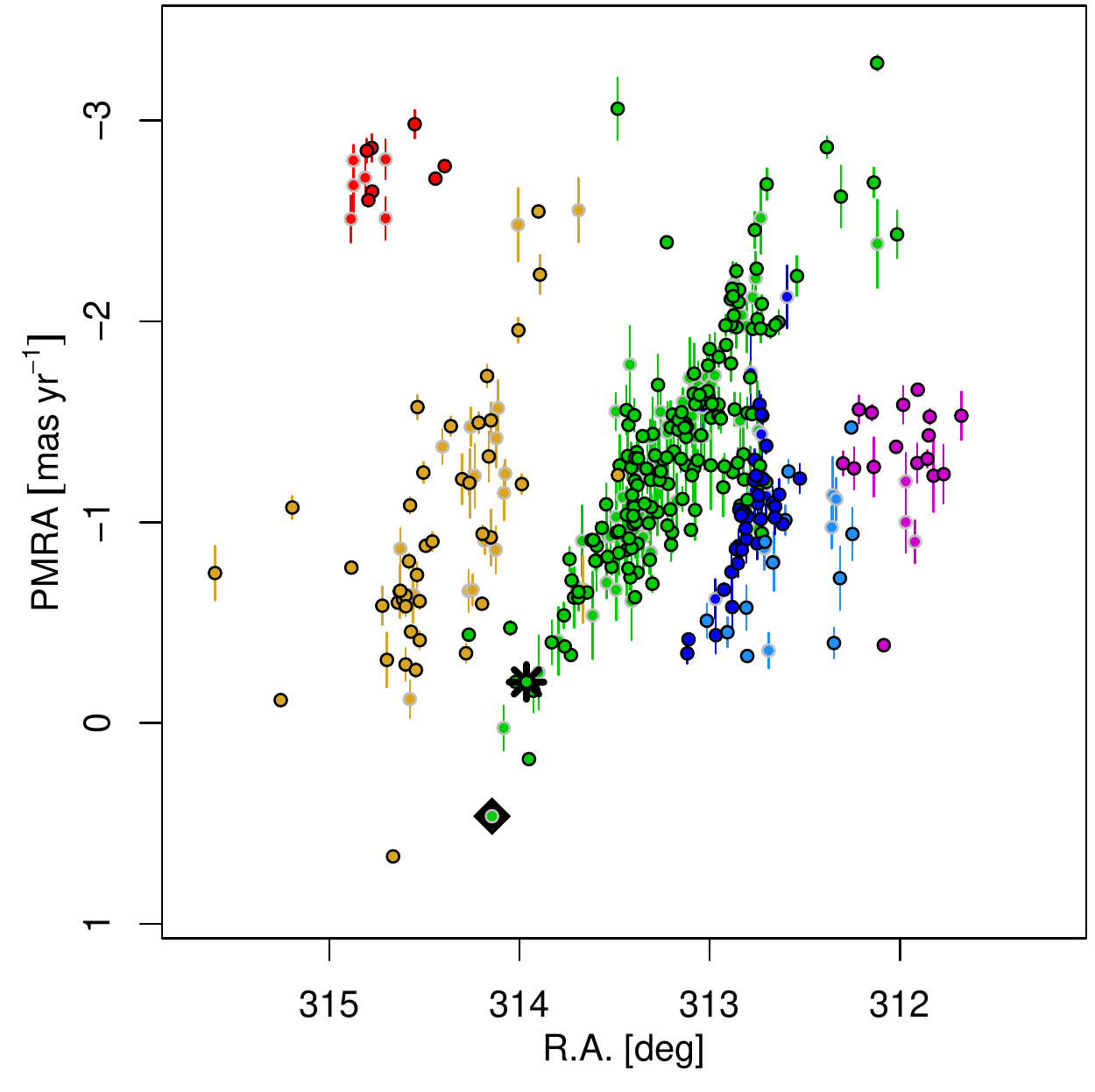}
\includegraphics[width=0.95\textwidth]{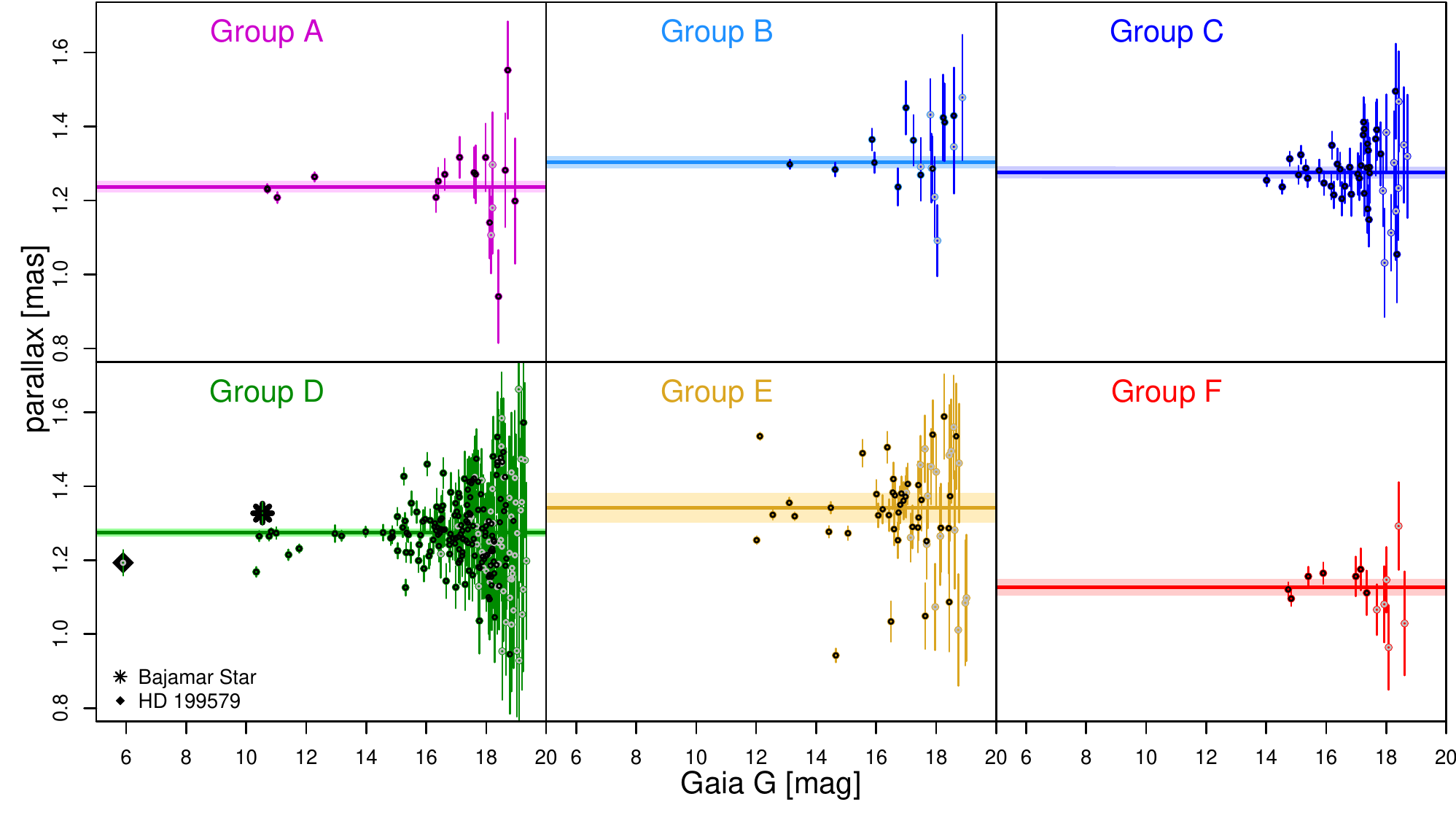}
\caption{Distributions of Gaia EDR3 parallaxes, proper motions, and positions for stars in the North America/Pelican Nebulae; the Bajamar Star (black asterisk) and HD~199579 (black diamond) are marked. Color coding indicates the groups from \citet{Kuhn2020}. Top left: Arrows indicate relative stellar motions after subtracting the median proper motion of the system. These are plotted on a map of the region, with optical nebulosity outlined in gray. Top right: Proper motion $\mu_{\alpha^\star}$ versus right ascension. Using EDR3, the groups become sharper than in \citet[][their Figure~5]{Kuhn2020}.  Bottom: Parallax versus $G$ magnitude for stars in each group. Stars are plotted with 1$\sigma$ error bars. The horizontal lines indicate the mean parallaxes, with 2$\sigma$ confidence intervals shaded. Sources outside the domain for the zero-point function have gray circles. With EDR3, the uncertainties on the means have decreased compared to DR2 \citep[][their Figure 7]{Kuhn2020}. Furthermore, the parallaxes for the O stars are now closer to the means than they were in DR2.   
\label{fig:1}}
\end{center}
\end{figure}

\citet{Kuhn2020} divided the NAP members into 6 subgroups (labeled A--F; Figure~\ref{fig:1}), which they found to have slightly different mean parallaxes using Gaia DR2. To estimate the mean parallaxes of these groups, we use the same statistical model (their Equation~1) but substitute the EDR3 parallaxes. For YSO Groups A--F, this yields $\varpi=1.237\pm0.008$, $1.304\pm0.009$, $1.276\pm0.008$, $1.274\pm0.006$, $1.341\pm0.020$, $1.127\pm0.011$~mas, respectively. Two of the larger groups, D (in the ``Atlantic'' cloud) and E (in the ``Gulf of Mexico'' cloud) are discrepant from each other at the 3$\sigma$ level, implying that Group~E is $\sim$40~pc nearer to us than Group~D. Furthermore, Group F is $\sim$100~pc farther than Group~D ($>$10$\sigma$). Similarly sized parallax differences were also detected in the DR2 data. Figure~\ref{fig:1} (bottom) shows how the individual stellar parallaxes compare to the mean parallaxes of the groups.

In projection, the Bajamar Star lies between Groups D and E, and it was tentatively assigned to Group D by \citet{Kuhn2020} on the basis that its relative proper motion is directed away from the center of D and toward E. Its EDR3 parallax is slightly smaller than that of E and slightly larger than that of D. This is an improvement over the situation with DR2 because we expect this star to be located behind Group E's ``Gulf of Mexico'' cloud. However, the formal uncertainty on the Bajamar Star's parallax is large enough that it cannot be determined which of these two groups it is closer to.

Given that Group~D appears to be relatively central in the NAP region, we use its parallax to calculate the distance of the complex. When interpreting this parallax as a distance, we take into account the  $\pm$0.026~mas spatially correlated systematic parallax uncertainty reported by \citet[][Section~5.6]{Lindegren2020_as}. Thus, the distance to the center of the NAP region is 785$\pm$16~pc. 

The median proper motion of the system is $(\mu_{\alpha^\star},\mu_\delta) = (-1.22,-3.27)$ in mas~yr$^{-1}$, with uncertainty  dominated by spatially correlated systematic errors \citep[$\pm$0.023~mas~yr$^{-1}$;][]{Lindegren2020_as}. The relative proper motions of the NAP stars, after subtracting the system's median motion, are indicated by arrows on a map of the region in Figure~\ref{fig:1} (top left).  
Figure~\ref{fig:1} (top right) shows position versus proper motion in right ascension for these stars. The Bajamar Star is marked on both diagrams, and it is clear that in EDR3 it is still an outlier in velocity. 

Using Equations~1--4 from \citet{Kuhn2019} to convert proper motions to projected velocities, with first-order corrections for perspective expansion and coordinate effects, this corresponds to a relative tangential velocity of $v=6.0\pm0.2$~km~s$^{-1}$ for the Bajamar Star. This is consistent with the Bajamar Star having escaped from a subgroup at a ``walkaway'' velocity \citep[][]{Schoettler2019}.

At the edge of the H\,{\sc ii} region there is a second O star, HD~199579 (also marked in Figure~\ref{fig:1}), with a corrected EDR3 parallax of 1.193$\pm$0.035. (Its magnitude, $G=5.9$~mag, is slightly brighter than the limit for the zero-point function, so $G=6$~mag is assumed.) This parallax suggests that the star is more distant than the main complex, but the parallax is only discrepant from Group~D by $\sim$2$\sigma$, which could arise due to statistical measurement error. The relative tangential velocity is 7.3$\pm$0.3~km~s$^{-1}$, making it another possible ``walkaway.'' 

In summary, we have reconsidered the kinematics of two O-type stars projected near the NAP region,
finding new compelling evidence for association of the Bajamar Star with the region,
but only marginal evidence for HD~199579 being associated.  
Gaia EDR3 data further substantiate the conclusions regarding these two stars in \citet{Kuhn2020}.

\acknowledgments This work is based on data from ESA's Gaia mission \citep{Gaia2016}, processed by the Data Processing and Analysis Consortium, funded by national institutions, particularly those participating in the Gaia Multilateral Agreement.


\begin{thebibliography}{}

\bibitem[Bally \& Scoville(1980)]{Bally1980} Bally, J. \& Scoville, N.~Z.\ 1980, \apj, 239, 121

\bibitem[Comer{\'o}n \& Pasquali(2005)]{Comeron2005} Comer{\'o}n, F. \& Pasquali, A.\ 2005, \aap, 430, 541

\bibitem[Dale(2015)]{Dale2015} Dale, J.~E.\ 2015, \nar, 68, 1

\bibitem[Fang et al.(2020)]{Fang2020} Fang, M., Hillenbrand, L.~A., Kim, J.~S., et al.\ 2020, \apj, 904, 146

\bibitem[Gaia Collaboration(2020)]{Gaia2020} Gaia Collaboration, Brown, A.\ G.\ A., Vallenari, A., et al.\ 2020, \aap, in press, arXiv:2012.01533

\bibitem[Gaia Collaboration et al.(2016)]{Gaia2016} Gaia Collaboration, Prusti, T., de Bruijne, J.~H.~J., et al.\ 2016, \aap, 595, A1

\bibitem[Kuhn et al.(2020)]{Kuhn2020} Kuhn, M.~A., Hillenbrand, L.~A., Carpenter, J.~M., et al.\ 2020, \apj, 899, 128

\bibitem[Kuhn et al.(2019)]{Kuhn2019} Kuhn, M.~A., Hillenbrand, L.~A., Sills, A., et al.\ 2019, \apj, 870, 32

\bibitem[Lindegren(2020a)]{Lindegren2020_zp}  Lindegren, L., Bastian, U., Biermann, M., et al.\ 2020, \aap, submitted, arXiv:2012.01742 

\bibitem[Lindegren(2020b)]{Lindegren2020_as} Lindegren, L.,  Klioner, S.\ A., Hern\'andez, J., et al.\ 2020, \aap, in press, arXiv:2012.03380

\bibitem[Ma{\'\i}z Apell{\'a}niz et al.(2016)]{MA2016} Ma{\'\i}z Apell{\'a}niz, J., Sota, A., Arias, J.~I., et al.\ 2016, \apjs, 224, 4

\bibitem[Schoettler et al.(2019)]{Schoettler2019} Schoettler, C., Parker, R.~J., Arnold, B., et al.\ 2019, \mnras, 487, 4615

\bibitem[Oh \& Kroupa(2016)]{Oh2016} Oh, S. \& Kroupa, P.\ 2016, \aap, 590, A107

\bibitem[Zucker et al.(2020)]{Zucker2020} Zucker, C., Speagle, J.~S., Schlafly, E.~F., et al.\ 2020, \aap, 633, A51

\end{thebibliography}
\end{document}